\documentclass[12pt]{article}
\usepackage{epsfig,amsmath,amsfonts,amssymb,amstext,afterpage,psfrag,
}
\usepackage{slashed}
\usepackage[square,comma,sort&compress,numbers]{natbib}
\allowdisplaybreaks
\newdimen\TW
\usepackage{color}
\definecolor{light-gray}{gray}{0.9}
\definecolor{dark-gray}{gray}{0.7}
\usepackage{tikz}

\long\def\symbolfootnote[#1]#2{\begingroup%
\def\thefootnote{\fnsymbol{footnote}}\footnote[#1]{#2}\endgroup}

\def\l{\langle}
\def\r{\rangle}

\def\spose#1{\hbox to 0pt{#1\hss}}
\def\lsim{\mathrel{\spose{\lower 3pt\hbox{$\mathchar"218$}}
 \raise 2.0pt\hbox{$\mathchar"13C$}}}
\def\gsim{\mathrel{\spose{\lower 3pt\hbox{$\mathchar"218$}}
 \raise 2.0pt\hbox{$\mathchar"13E$}}}

\catcode`@=11
\def\@citex[#1]#2{%
  \if@filesw\immediate\write\@auxout{\string\citation{#2}}\fi
  \def\@citea{}\@cite{\@for\@citeb:=#2\do
    {\@citea\def\@citea{,\penalty\@m}\@ifundefined
      {b@\@citeb}{{\bf ?}\@warning
{Citation `\@citeb' on page \thepage \space undefined}}%
      \hbox{\csname b@\@citeb\endcsname}}}{#1}}
\def\citer{\@ifnextchar [{\@tempswatrue\@citexr}{\@tempswafalse\@citexr[]}}
  \def\@citexr[#1]#2{%
    \if@filesw\immediate\write\@auxout{\string\citation{#2}}\fi
    \def\@citea{}\@cite{\@for\@citeb:=#2\do
      {\@citea\def\@citea{--\penalty\@m}\@ifundefined
{b@\@citeb}{{\bf ?}\@warning
{Citation `\@citeb' on page \thepage \space undefined}}%
\hbox{\csname b@\@citeb\endcsname}}}{#1}}

\begin{document}

\begin{titlepage}

\begin{flushright}
{\small
LMU-ASC~19/15\\ 
April 2015\\
}
\end{flushright}

\vspace{0.5cm}
\begin{center}
{\Large\bf \boldmath                                               
Note on Anomalous Higgs-Boson Couplings\\
\vspace*{0.3cm}
in Effective Field Theory
\unboldmath}
\end{center}

\vspace{0.5cm}
\begin{center}
{\sc G. Buchalla, O. Cat\`a, A. Celis and C. Krause} 
\end{center}

\vspace*{0.4cm}

\begin{center}
Ludwig-Maximilians-Universit\"at M\"unchen, Fakult\"at f\"ur Physik,\\
Arnold Sommerfeld Center for Theoretical Physics, 
D--80333 M\"unchen, Germany
\end{center}

\vspace{1.5cm}
\begin{abstract}
\vspace{0.2cm}\noindent
We propose a parametrization of anomalous Higgs-boson 
couplings that is both systematic and practical.
It is based on the electroweak chiral Lagrangian, including a
light Higgs boson, as the effective field theory (EFT) at the
electroweak scale $v$. This is the appropriate framework for the case
of sizeable deviations in the Higgs couplings of order $10\%$ from the
Standard Model, considered to be parametrically larger than new-physics
effects in the sector of electroweak gauge interactions.
The role of power counting in identifying the relevant parameters
is emphasized. 
The three relevant scales, $v$, the scale of new Higgs dynamics $f$, 
and the cut-off $\Lambda=4\pi f$, admit expansions in $\xi=v^2/f^2$
and $f^2/\Lambda^2$. The former corresponds to an organization of 
operators by their canonical dimension, the latter by their
loop order or chiral dimension. In full generality the EFT is thus 
organized as a double expansion. However, as long as $\xi\gg 1/16\pi^2$ 
the EFT systematics is closer to the chiral counting.
The leading effects in the consistent approximation provided by the
EFT, relevant for the presently most important processes of Higgs 
production and decay, are given by a few (typically six) couplings.
These parameters allow us to describe the properties of the Higgs boson
in a general and systematic way, and with a precision adequate for the
measurements to be performed at the LHC. 
The framework can be systematically extended to include loop corrections
and higher-order terms in the EFT.

\end{abstract}

\vfill

\end{titlepage}

\section{Introduction}
\label{sec:intro}
The discovery of the Higgs boson at the Large Hadron Collider (LHC)
\cite{Aad:2012tfa}
has focused current research in high-energy physics onto the 
detailed investigation of its properties. Observing deviations 
from the predictions of the Standard Model (SM) would give 
us important information on the dynamics of electroweak symmetry breaking.
The question of how to obtain an efficient parametrization of Higgs couplings
is under active discussion at present (see \cite{David}
and {\it e.g.} \cite{Gonzalez-Alonso:2014eva} for a specific proposal).

Assuming a mass gap, with new degrees of freedom not much below the TeV scale,
a general and model-independent parametrization of new physics can be 
achieved within the framework of an effective field theory (EFT).
The EFT as the low-energy approximation of new physics at high energies
(`bottom-up' perspective) needs to be defined by its
{\it particle content}, the relevant {\it symmetries}, and an appropriate
{\it power counting}.

At present, data on Higgs-boson couplings \cite{Khachatryan:2014jba}
still allow deviations from
the Standard Model of order $10\%$, much larger than in the sector of
the usual electroweak precision tests with gauge bosons.
This leads one to consider the interesting scenario, relevant for
Higgs studies at the LHC, in which non-standard contributions to Higgs 
couplings are indeed of this size. Such effects would point to a new-physics
scale $f$ of typically $500$ -- $1000\,{\rm GeV}$, corresponding to
deviations characterized by the parameter $\xi\equiv v^2/f^2={\cal O}(10\%)$,
where $v=246\, {\rm GeV}$ is the electroweak scale.
Examples for such new dynamics in the Higgs sector at scale $f$ are given, 
in particular, by models with a composite, pseudo-Goldstone Higgs particle 
\cite{Agashe:2004rs,Contino:2006qr,Contino:2010rs,Falkowski:2007hz,Carena:2014ria}, but also by other models with a modified Higgs sector at either weak or 
strong coupling.

Anomalous contributions, with respect to the Standard Model, of order $\xi$
in the Higgs couplings will generically lead to a cut-off $\Lambda=4\pi f$
in the effective description of the new Higgs dynamics. This picture might 
be supplemented by TeV-scale (order $f$) new degrees of freedom
(non-standard fermions, extra pseudo-Goldstone bosons), understood to be
integrated out in the EFT at the electroweak scale $v$.

As has been discussed in \cite{Buchalla:2014eca}, 
the EFT can then be organized in full generality as a double expansion
in $\xi=v^2/f^2$ and $f^2/\Lambda^2=1/16\pi^2$, which are the
two dimensionless parameters that can be formed out of the three
relevant scales $v$, $f$ and $\Lambda$. They are both small under
the condition $v\ll f\ll \Lambda$.
The expansion in $\xi$ amounts to an expansion of the Lagrangian in 
operators of increasing {\it canonical dimension} ($d$). The expansion in
$f^2/\Lambda^2$ corresponds to a loop expansion or, equivalently,
to an expansion in terms of increasing {\it chiral dimension} ($\chi$).
For the phenomenologically interesting case where $\xi \gg f^2/\Lambda^2$,
the character of the expansion is dominated by chiral counting rather
than by canonical dimensions. It is therefore convenient to phrase
the effective theory from the outset in terms of the nonlinear
electroweak chiral Lagrangian.\footnote{The chiral Lagrangian for the
(Higgs-less) electroweak Standard Model has first been developed 
in \cite{Appelquist:1980vg}. The extension with a light Higgs boson
has been treated in  
\cite{Feruglio:1992wf,Buchalla:2013eza,Buchalla:2012qq,Buchalla:2013rka}.
A complete presentation of power counting and next-to-leading order
terms has been given in
\cite{Buchalla:2013eza,Buchalla:2012qq,Buchalla:2013rka}.}
This automatically implies a resummation to all orders in $\xi$.

The interesting feature of parametrically larger new-physics effects in the 
Higgs sector ($\sim 1/f^2$) than in the gauge sector ($\sim 1/\Lambda^2$),
in the context of composite-Higgs scenarios, has been pointed out in
\cite{Giudice:2007fh,Contino:2013kra}. However, the EFT formulation 
discussed there (SILH Lagrangian) follows a dimensional counting 
to describe the leading effects, which is not fully adequate in 
this case \cite{Buchalla:2014eca}.

Under the assumptions stated above, the leading new physics effects 
in the Higgs sector, of order $\xi$, will be essentially described by the 
leading-order chiral Lagrangian, with qualifications to be discussed below.
As a consequence, most effects from the next-to-leading order chiral
Lagrangian, of order $\xi/16\pi^2=v^2/\Lambda^2$ can be consistently
neglected. This will result in a considerable reduction of the
number of parameters, while still accounting for the dominant effects
of new dynamics in the Higgs sector in a general and systematic way.

Similar parametrizations based on the leading-order chiral Lagrangian
have been considered and employed before by many authors (see
\cite{Azatov:2012bz}
and references therein). 
The essential new aspect of our discussion is that it is based
on a general and consistent power counting and the consideration
of the next-to-leading order chiral Lagrangian in assessing
the size of subleading corrections. 

The remainder of this note is organized as follows.
Section \ref{sec:leff} summarizes the effective Lagrangian
and the underlying assumptions. Section \ref{sec:param} introduces
our parametrization of anomalous Higgs couplings and outlines
strategies for phenomenological applications. We conclude in 
Section \ref{sec:concl}.

\section{Effective Lagrangian}
\label{sec:leff}

The most important assumptions that define the EFT of 
new physics in the Higgs sector based on the electroweak chiral Lagrangian
can be summarized as follows:
\begin{description}
\item[(i)]
SM particle content
\item[(ii)]
symmetries:
\begin{itemize}
\item
SM gauge symmetries
\item
conservation of lepton and baryon number 
\item 
conservation at lowest order of custodial symmetry, CP invariance
in the Higgs sector, lepton flavour
\end{itemize}
The symmetries under the third item may be violated at some level, but this 
would only affect terms at subleading order.
We consider these assumptions as affecting the generality
of the EFT only very mildly. Generalizations may in principle be introduced
if necessary.
\item[(iii)]
power counting by chiral dimensions (loop expansion):\\
The loop expansion is equivalent to the counting of chiral dimensions
\cite{Buchalla:2013eza}, with the simple assignment
\begin{itemize}
\item
$0$ for bosons (gauge fields, Goldstones and Higgs)
\item
$1$ for each derivative, weak coupling (e.g. gauge or Yukawa),
and fermion bilinear
\end{itemize}
The loop order $L$ of a term in the Lagrangian is equivalent
to its chiral dimension (or chiral order) $2L+2$.
We note that the loop expansion is not equivalent to a pure
derivative counting in the presence of gauge interactions and fermions.
\end{description}

To leading order in chiral dimensions ($\chi=2$)  
the effective Lagrangian can then be written as \cite{Buchalla:2013rka} 
\begin{eqnarray}\label{l2}
{\cal L}_2 &=& -\frac{1}{2} \langle G_{\mu\nu}G^{\mu\nu}\rangle
-\frac{1}{2}\langle W_{\mu\nu}W^{\mu\nu}\rangle 
-\frac{1}{4} B_{\mu\nu}B^{\mu\nu}
+\bar q i\!\not\!\! Dq +\bar l i\!\not\!\! Dl
 +\bar u i\!\not\!\! Du +\bar d i\!\not\!\! Dd +\bar e i\!\not\!\! De 
\nonumber\\
&& +\frac{v^2}{4}\ \l L_\mu L^\mu \r\, \left( 1+F_U(h)\right)
+\frac{1}{2} \partial_\mu h \partial^\mu h - V(h) \nonumber\\
&& - v \left[ \bar q \left( Y_u +
       \sum^\infty_{n=1} Y^{(n)}_u \left(\frac{h}{v}\right)^n \right) U P_+r 
+ \bar q \left( Y_d + 
     \sum^\infty_{n=1} Y^{(n)}_d \left(\frac{h}{v}\right)^n \right) U P_-r
  \right. \nonumber\\ 
&& \quad\quad\left. + \bar l \left( Y_e +
   \sum^\infty_{n=1} Y^{(n)}_e \left(\frac{h}{v}\right)^n \right) U P_-\eta 
+ {\rm h.c.}\right]
\end{eqnarray}
with $L_\mu=i UD_\mu U^\dagger$, $P_\pm = 1/2\pm T_3$.
Here $F_U=\sum^\infty_{n=1} f_{U,n} (h/v)^n$, $V=v^4\sum^\infty_{n=2} f_{V,n}(h/v)^n$.

In addition to the leading-order terms, the Higgs-photon-photon and
Higgs-gluon-gluon couplings, of the form
\begin{equation}
h F^{\mu\nu}F_{\mu\nu},\qquad h G^{\mu\nu}G_{\mu\nu}
\end{equation}
from the Lagrangian of chiral dimension 4
have to be included to account for all the leading effects. This is
because these couplings are loop-induced in the Standard Model and the
relative corrections at chiral dimension 4 are also of order $\xi$.
In contrast, the corresponding terms
\begin{equation}\label{hwwnlo}
h W^{+\mu\nu}W^-_{\mu\nu},\qquad h Z^{\mu\nu}Z_{\mu\nu}
\end{equation}
are subleading compared to the dominant $hWW$ and $hZZ$ couplings
from (\ref{l2}) and can be neglected.
The same is true, in particular, for modifications of the gauge-fermion
couplings, which also arise at chiral dimension 4. Focussing
on the modified Higgs coupling in $h\to ZZ^*\to 4l$, for instance,
and assuming Standard-Model couplings for $Z\to ll$, is therefore
a consistent approximation.

If custodial symmetry is only broken by weak perturbations every spurion that 
breaks this symmetry will come with a chiral dimension and operators breaking 
custodial symmetry will be further suppressed.

\section{Parametrization of Higgs Couplings}
\label{sec:param}

Based on the discussion of power counting in the previous section we can now 
define the parametrization of the Higgs couplings in a systematic way. 
With the foreseeable precision of the data at the LHC, we are predominantly 
sensitive
to leading deviations from the SM. The main input given by the 
experiments are the signal strengths $\mu$. We will therefore start from 
$\mu$ and consider the leading deviations given by the power counting of 
the EFT. The signal strength is defined as
\begin{equation}
  \label{eq:3.1}
  \mu = \frac{\sigma(X)\cdot BR(h\rightarrow Y)}{\sigma(X)_{\text{SM}}\cdot 
            BR(h\rightarrow Y)_{\text{SM}}},
\end{equation}
where $\sigma(X)$ denotes the production cross section of the Higgs in the 
process $X$ and $BR(h\rightarrow Y)$ is the branching ratio of Higgs decaying 
to the final state $Y$. Possible processes in the production are gluon fusion, 
Higgs-strahlung from vector bosons, vector boson fusion and $\bar{t}t$ fusion: 
$X\in \{ ggH, WH/ZH, VBF, ttH\}$. The relevant decay channels are 
Higgs to bottom quark pairs, tau leptons, as well as $W$, $Z$ and photon pairs: 
$Y\in \{ bb, \tau\tau, WW, ZZ, \gamma\gamma\}$. \\
These Standard Model processes already fall in two categories: tree and 
loop-level induced processes. The tree-level processes can be affected by the 
leading order Lagrangian and power counting tells us that deviations of order 
$\mathcal{O}(\xi)$ might be expected. The loop-induced processes 
($ggH$, $\gamma\gamma H$) are suppressed 
by a factor of $1/16\pi^{2}$ with respect to the tree level ones. However, 
there are local terms at next-to-leading order (chiral order 4) in the 
Lagrangian with size of order $\mathcal{O}(\xi/ 16\pi^{2})$. This  
means that such corrections  are of $\mathcal{O}(\xi)$ relative to the SM 
and have to be kept as well. 
The Lagrangian that results from these considerations is given by
\begin{equation}
  \label{eq:3.2}
\begin{array}{ll}
  \mathcal{L} &=2 c_{V} \left(m_{W}^{2}W_{\mu}^{+}W^{-\mu} 
+\frac{1}{2} m^2_Z Z_{\mu}Z^{\mu}\right) \dfrac{h}{v} -c_{t} y_{t} \bar{t} t h 
- c_{b} y_{b} \bar{b}b h -c_{\tau} y_{\tau} \bar{\tau}\tau h  \\
 &+ \dfrac{e^{2}}{16\pi^{2}} c_{\gamma\gamma} F_{\mu\nu}F^{\mu\nu} \dfrac{h}{v}  
+\dfrac{g_{s}^{2}}{16\pi^{2}} c_{gg}\langle G_{\mu\nu}G^{\mu\nu}\rangle\dfrac{h}{v}
\end{array}
\end{equation}
where $y_f=m_f/v$.
The SM at tree level is given by $c_{V}= c_{t} = c_{b}=c_{\tau} =1$ and 
$c_{gg} =c_{\gamma\gamma} =0$. Deviations due to new physics are expected to 
start at $\mathcal{O}(\xi)$.
The couplings written in (\ref{eq:3.2}) are obtained from the effective
Lagrangian in (\ref{l2}) by extracting the terms with a single field $h$, 
in unitary gauge for the $W$ and $Z$, and neglecting (small) flavour
violation and light fermions. According to the assumptions stated in 
Section \ref{sec:leff} custodial symmetry holds at leading order, 
implying $c_W\equiv c_Z\equiv c_V$. In addition, the local terms with 
$c_{\gamma\gamma}$ and $c_{gg}$ have to be added.
 
The terms that we may neglect in our analysis fall into two groups: i) Terms at 
the same (chiral) order of the EFT but with a numerically very small impact 
on current observables. Examples for this group are the 
$hZ\gamma$ local operator and the coupling to light fermions. ii) Terms 
of higher order in the chiral  
expansion, which can be neglected based on the EFT power counting.
Operators of this type are for example the NLO contributions to 
$hW^{+}W^{-}$ and $hZZ$ in (\ref{hwwnlo}). 

The fact that operators of group ii) can be neglected is illustrated
{\it e.g.} by the analysis of the contributions of NLO operators to 
$h\to Z \ell^+\ell^-$ \cite{Buchalla:2013mpa}.
In processes with off-shell Higgs production, the energy relevant for the
higher-derivative Higgs couplings can become numerically larger than the 
scale $v$ and lead to some enhancement of these corrections. However, 
as for any EFT, the size of higher-order corrections is required to be 
sufficiently small in order to ensure the validity of the EFT expansion.

The discussion above shows that the 
$\kappa$-formalism \cite{Heinemeyer:2013tqa} that is widely used as a simple 
approximation is indeed not only motivated phenomenologically but can be
justified from an effective field theory. The first deviations from the SM are 
expected to be in the event rates. Deviations in the shapes are 
subleading compared to them. However, there is one main difference between 
the approach presented here and the $\kappa$-formalism described in 
\cite{Heinemeyer:2013tqa}. The $\kappa_{i}$'s for the one-loop processes of 
Higgs coupling to a pair of photons/gluons are either given as a function 
of the modified couplings of Higgs to vectors/fermions or as a free parameter 
describing in addition possible new particles in the loop as a point 
interaction. Our approach takes both of these possibilities separately into 
account. Even though the number of free parameters is not changed, 
this makes the interpretation of the results more transparent. 

A numerical analysis of the currently accessible Higgs channels
within the framework described above will be presented elsewhere \cite{BCCK}.
Such an analysis could further be extended to additional processes,
for instance $h\to Z\gamma$ decay or double-Higgs production.

Once the experimental precision improves to the (sub)percent level in
the Higgs couplings, the analysis outlined above has to be generalized 
beyond the lowest order. 
This can be done in a systematic way by considering the two groups 
of operators that have been neglected in a first approximation.

\section{Conclusions}
\label{sec:concl}

The upcoming run of the LHC has the potential to detect anomalous
Higgs-boson couplings, where new-physics effects may still be of
order $10\%$, considerably larger than in the well-tested electroweak
gauge sector. In this note we have put forward a formalism able to describe
these potential new-physics effects in a model-independent way.

\begin{itemize}
\item
We have argued that the electroweak chiral Lagrangian, including a light
Higgs boson, is a suitable framework to test such a scenario, which
is particularly relevant phenomenologically during the coming years.

\item
The chiral Lagrangian, being an effective field theory, comes naturally with a
power counting that allows for well-defined approximations and for systematic
improvements including higher-order corrections.

\item
The leading-order chiral Lagrangian, which precisely captures the
potentially sizable new-physics effects in the Higgs sector, can be
used as a first, well-defined approximation. It has the practical benefit 
of reducing the number of relevant parameters to a manageable set. 
With slight modifications, it actually amounts to an effective-field theory 
justification of the usual $\kappa$-formalism.

\item
The chiral Lagrangian is based on a loop expansion or, equivalently,
a power counting in terms of chiral dimensions ($0$ for bosons, $1$
for derivatives, weak couplings and fermion bilinears). This is in
contrast to the more common counting in terms of
canonical dimension, which implies a different ordering of operators
in the effective theory that does not naturally single out the new-physics
effects in the Higgs sector as the dominant ones.
\end{itemize}

While the chiral Lagrangian description has been used before in 
phenomenological applications, we have emphasized here the role of chiral 
counting in establishing a parametrization of anomalous Higgs couplings.

\section*{Acknowledgements}

This work was performed 
in the context of the ERC Advanced Grant project `FLAVOUR' (267104) and was 
supported in part by the DFG cluster of excellence `Origin and Structure 
of the Universe' and DFG grant BU 1391/2-1. A.C. is supported
by a Research Fellowship of the Alexander von Humboldt Foundation.



\begin{thebibliography}{99}

\bibitem{Aad:2012tfa} 
  G.~Aad {\it et al.}  [ATLAS Collaboration],
  Phys.\ Lett.\ B {\bf 716}, 1 (2012)
  [arXiv:1207.7214 [hep-ex]];
  S.~Chatrchyan {\it et al.}  [CMS Collaboration],
  Phys.\ Lett.\ B {\bf 716}, 30 (2012)
  [arXiv:1207.7235 [hep-ex]];
  ATLAS Collaboration,
  ATLAS-CONF-2013-034;
  CMS Collaboration,
  CMS-PAS-HIG-13-005;
  G.~Aad {\it et al.}  [ ATLAS Collaboration],
  arXiv:1307.1432 [hep-ex].

\bibitem{David}
A.~David, {\it A middle way between kappas and Wilson coefficients},
talk at Moriond EW 2015, https://indico.in2p3.fr/event/10819/

\bibitem{Gonzalez-Alonso:2014eva}
  M.~Gonzalez-Alonso, A.~Greljo, G.~Isidori and D.~Marzocca,
  Eur.\ Phys.\ J.\ C {\bf 75} (2015) 3,  128
  [arXiv:1412.6038 [hep-ph]].

\bibitem{Khachatryan:2014jba}
  V.~Khachatryan {\it et al.}  [CMS Collaboration],
  arXiv:1412.8662 [hep-ex];
  The ATLAS collaboration,
  ATLAS-CONF-2015-007, ATLAS-COM-CONF-2015-011.

\bibitem{Agashe:2004rs}
  K.~Agashe, R.~Contino and A.~Pomarol,
  Nucl.\ Phys.\ B {\bf 719} (2005) 165
  [hep-ph/0412089].

\bibitem{Contino:2006qr}
  R.~Contino, L.~Da Rold and A.~Pomarol,
  Phys.\ Rev.\ D {\bf 75} (2007) 055014
  [hep-ph/0612048].

\bibitem{Contino:2010rs}
  R.~Contino,
  arXiv:1005.4269 [hep-ph].

\bibitem{Falkowski:2007hz} 
  A.~Falkowski,
  Phys.\ Rev.\ D {\bf 77}, 055018 (2008)
  [arXiv:0711.0828 [hep-ph]].

\bibitem{Carena:2014ria} 
  M.~Carena, L.~Da Rold and E.~Pont\'on,
  JHEP {\bf 1406}, 159 (2014)
  [arXiv:1402.2987 [hep-ph]].


\bibitem{Buchalla:2014eca} 
  G.~Buchalla, O.~Cat\`a and C.~Krause,
  Nucl.\ Phys.\ B {\bf 894}, 602 (2015)
  [arXiv:1412.6356 [hep-ph]].

\bibitem{Appelquist:1980vg}
  T.~Appelquist and C.~W.~Bernard,
  Phys.\ Rev.\  D {\bf 22}, 200 (1980).
  A.~C.~Longhitano,
  Phys.\ Rev.\  D {\bf 22}, 1166 (1980).
  T.~Appelquist, M.~J.~Bowick, E.~Cohler and A.~I.~Hauser,
  Phys.\ Rev.\  D {\bf 31}, 1676 (1985).

\bibitem{Feruglio:1992wf}
  F.~Feruglio,
  Int.\ J.\ Mod.\ Phys.\ A {\bf 8}, 4937 (1993)
  [hep-ph/9301281];
  J.~Bagger {\it et al.},
  Phys.\ Rev.\ D {\bf 49}, 1246 (1994)
  [hep-ph/9306256];
  V.~Koulovassilopoulos and R.~S.~Chivukula,
  Phys.\ Rev.\ D {\bf 50}, 3218 (1994)
  [hep-ph/9312317];
  C.~P.~Burgess, J.~Matias and M.~Pospelov,
  Int.\ J.\ Mod.\ Phys.\ A {\bf 17}, 1841 (2002)
  [hep-ph/9912459];
  L.~M.~Wang and Q.~Wang,
  Chin.\ Phys.\ Lett.\  {\bf 25}, 1984 (2008)
  [hep-ph/0605104].
  B.~Grinstein and M.~Trott,
  Phys.\ Rev.\ D {\bf 76}, 073002 (2007)
  [arXiv:0704.1505 [hep-ph]].
  R.~Alonso, M.~B.~Gavela, L.~Merlo, S.~Rigolin and J.~Yepes,
  Phys.\ Lett.\ B {\bf 722}, 330 (2013)
  [Erratum-ibid.\ B {\bf 726}, 926 (2013)]
  [arXiv:1212.3305 [hep-ph]].

\bibitem{Buchalla:2013eza}
  G.~Buchalla, O.~Cat\`a and C.~Krause,
  Phys.\ Lett.\ B {\bf 731}, 80 (2014)
  [arXiv:1312.5624 [hep-ph]].

\bibitem{Buchalla:2012qq} 
  G.~Buchalla and O.~Cat\`a,
  JHEP {\bf 1207}, 101 (2012)
  [arXiv:1203.6510 [hep-ph]].

\bibitem{Buchalla:2013rka} 
  G.~Buchalla, O.~Cat\`a and C.~Krause,
  Nucl.\ Phys.\ B {\bf 880}, 552 (2014)
  [arXiv:1307.5017 [hep-ph]].

\bibitem{Giudice:2007fh} 
  G.~F.~Giudice, C.~Grojean, A.~Pomarol and R.~Rattazzi,
  JHEP {\bf 0706}, 045 (2007)
  [hep-ph/0703164].

\bibitem{Contino:2013kra} 
  R.~Contino, M.~Ghezzi, C.~Grojean, M.~M\"uhlleitner and M.~Spira,
  JHEP {\bf 1307}, 035 (2013)
  [arXiv:1303.3876 [hep-ph]].

\bibitem{Azatov:2012bz}
  D.~Carmi, A.~Falkowski, E.~Kuflik and T.~Volansky,
  JHEP {\bf 1207}, 136 (2012)
  [arXiv:1202.3144 [hep-ph]];
  A.~Azatov, R.~Contino and J.~Galloway,
  JHEP {\bf 1204} (2012) 127
   [Erratum-ibid.\  {\bf 1304} (2013) 140]
  [arXiv:1202.3415 [hep-ph]];
  J.~R.~Espinosa, C.~Grojean, M.~M\"uhlleitner and M.~Trott,
  JHEP {\bf 1205}, 097 (2012)
  [arXiv:1202.3697 [hep-ph]];
  A.~Azatov {\it et al.}, 
  JHEP {\bf 1206} (2012) 134
  [arXiv:1204.4817 [hep-ph]];
  J.~R.~Espinosa, C.~Grojean, M.~M\"uhlleitner and M.~Trott,
  JHEP {\bf 1212} (2012) 045
  [arXiv:1207.1717 [hep-ph]];
  P.~P.~Giardino {\it et al.}, 
  JHEP {\bf 1405} (2014) 046
  [arXiv:1303.3570 [hep-ph]];
  J.~Ellis and T.~You,
  JHEP {\bf 1306} (2013) 103
  [arXiv:1303.3879 [hep-ph]].

\bibitem{Buchalla:2013mpa}
  G.~Buchalla, O.~Cat\`a and G.~D'Ambrosio,
  Eur.\ Phys.\ J.\ C {\bf 74} (2014) 3,  2798
  [arXiv:1310.2574 [hep-ph]].

\bibitem{Heinemeyer:2013tqa}
  S.~Heinemeyer {\it et al.}  [LHC Higgs Cross Section Working Group 
  Collaboration],
  arXiv:1307.1347 [hep-ph].

\bibitem{BCCK} G. Buchalla, O. Cat\`a, A. Celis and C. Krause,
  {\it in preparation.}

\end{thebibliography}
\end{document}